\documentstyle[12pt]{article}

\catcode`\@=11
\long\def\@makefntext#1{
\protect\noindent \hbox to 3.2pt {\hskip-.9pt
$^{{\ninerm\@thefnmark}}$\hfil}#1\hfill}		

 \def\@makefnmark{\hbox to 0pt{$^{\@thefnmark}$\hss}}  

\def\ps@myheadings{\let\@mkboth\@gobbletwo
\def\@oddhead{\hbox{}
\rightmark\hfil\ninerm\thepage}
\def\@oddfoot{}\def\@evenhead{\ninerm\thepage\hfil
\leftmark\hbox{}}\def\@evenfoot{}
\def\sectionmark##1{}\def\subsectionmark##1{}}

\newcounter{sectionc}\newcounter{subsectionc}\newcounter{subsubsectionc}
\renewcommand{\section}[1] {\vspace{0.6cm}\addtocounter{sectionc}{1}
\setcounter{subsectionc}{0}\setcounter{subsubsectionc}{0}\noindent
	{\bf\thesectionc. #1}\par\vspace{0.4cm}}
\renewcommand{\subsection}[1] {\vspace{0.6cm}\addtocounter{subsectionc}{1}
	\setcounter{subsubsectionc}{0}\noindent
	{\it\thesectionc.\thesubsectionc. #1}\par\vspace{0.4cm}}
\renewcommand{\subsubsection}[1]
{\vspace{0.6cm}\addtocounter{subsubsectionc}{1}
	\noindent {\rm\thesectionc.\thesubsectionc.\thesubsubsectionc.
	#1}\par\vspace{0.4cm}}

\newcounter{appendixc}
\newcounter{subappendixc}[appendixc]
\newcounter{subsubappendixc}[subappendixc]

\renewcommand{\appendix}[1] {\vspace{0.6cm}
        \refstepcounter{appendixc}
        \setcounter{figure}{0}
        \setcounter{table}{0}
        \setcounter{equation}{0}
        \renewcommand{\thefigure}{\Alph{appendixc}.\arabic{figure}}
        \renewcommand{\thetable}{\Alph{appendixc}.\arabic{table}}
        \renewcommand{\theappendixc}{\Alph{appendixc}}
        \renewcommand{\theequation}{\Alph{appendixc}.\arabic{equation}}
        \noindent{\bf Appendix \theappendixc #1}\par\vspace{0.4cm}}

\def\abstracts#1{{
	\centering{\begin{minipage}{30pc}\tenrm\baselineskip=12pt\noindent
	\centerline{\tenrm ABSTRACT}\vspace{0.3cm}
	\parindent=0pt #1
	\end{minipage}}\par}}

\renewenvironment{thebibliography}[1]
	{\begin{list}{\arabic{enumi}.}
	{\usecounter{enumi}\setlength{\parsep}{0pt}
\setlength{\leftmargin 1.25cm}{\rightmargin 0pt}
	 \setlength{\itemsep}{0pt} \settowidth
	{\labelwidth}{#1.}\sloppy}}{\end{list}}

\newcounter{itemlistc}
\newcounter{romanlistc}
\newcounter{alphlistc}
\newcounter{arabiclistc}

\newcommand{\fcaption}[1]{
        \refstepcounter{figure}
        \setbox\@tempboxa = \hbox{\tenrm Fig.~\thefigure. #1}
        \ifdim \wd\@tempboxa > 6in
           {\begin{center}
        \parbox{6in}{\tenrm\baselineskip=12pt Fig.~\thefigure. #1}
            \end{center}}
        \else
             {\begin{center}
             {\tenrm Fig.~\thefigure. #1}
              \end{center}}
        \fi}

\newcommand{\tcaption}[1]{
        \refstepcounter{table}
        \setbox\@tempboxa = \hbox{\tenrm Table~\thetable. #1}
        \ifdim \wd\@tempboxa > 6in
           {\begin{center}
        \parbox{6in}{\tenrm\baselineskip=12pt Table~\thetable. #1}
            \end{center}}
        \else
             {\begin{center}
             {\tenrm Table~\thetable. #1}
              \end{center}}
        \fi}

\def\@citex[#1]#2{\if@filesw\immediate\write\@auxout
	{\string\citation{#2}}\fi
\def\@citea{}\@cite{\@for\@citeb:=#2\do
	{\@citea\def\@citea{,}\@ifundefined
	{b@\@citeb}{{\bf ?}\@warning
	{Citation `\@citeb' on page \thepage \space undefined}}
	{\csname b@\@citeb\endcsname}}}{#1}}

\newif\if@cghi
\def\cite{\@cghitrue\@ifnextchar [{\@tempswatrue
	\@citex}{\@tempswafalse\@citex[]}}
\def\citelow{\@cghifalse\@ifnextchar [{\@tempswatrue
	\@citex}{\@tempswafalse\@citex[]}}
\def\@cite#1#2{{$\null^{#1}$\if@tempswa\typeout
	{IJCGA warning: optional citation argument
	ignored: `#2'} \fi}}

\def\fnt#1#2{\footnotetext{\kern-.3em
	{$^{\mbox{\sevenrm #1}}$}{#2}}}

 1
 1
 1

\font\tenbf=cmbx10
\font\tenrm=cmr10
\font\tenit=cmti10

\font\ninerm=cmr9

\newcommand{\beq}    {\begin{equation}}
\newcommand{\eeq}    {\end{equation}}
\newcommand{\beqarr} {\begin{eqnarray}}
\newcommand{\eeqarr} {\end{eqnarray}}
\newcommand{\barr}   {\begin{array}}
\newcommand{\earr}   {\end{array}}

\newcommand{\lsim}{\mathrel{\mathop{\kern 0pt \rlap
  {\raise.2ex\hbox{$<$}}}
  \lower.9ex\hbox{\kern-.190em $\sim$}}}
\newcommand{\gsim}{\mathrel{\mathop{\kern 0pt \rlap
  {\raise.2ex\hbox{$>$}}}
  \lower.9ex\hbox{\kern-.190em $\sim$}}}

\newcommand{\mb}[1]  {\mbox{#1}}
\newcommand{\mbi}[1] {\mbox{\scriptsize #1}}
\newcommand{\gev}    {\mb{GeV}}
\newcommand{\tev}    {\mb{TeV}}
\newcommand{\cm}     {\mb{cm}}
\newcommand{\second} {\mb{s}}
\newcommand{\gram}   {\mb{g}}
\newcommand{\br}     {\hfill\break}

\newcommand{\tb}     {\tan\beta}

\newcommand{\mchi}   {m_\chi}
\newcommand{\bi}     {{\tilde B}}
\newcommand{\wi}     {{\tilde W_3}}
\newcommand{\sigmav} {\sigma_{\mbi{ann}} v}

\begin{document}

\centerline{\tenbf SEARCHING FOR RELIC NEUTRALINOS}
\baselineskip=16pt
\vspace{-2.0cm}
\rightline{DFTT 67/94}
\rightline{GEF-Th-10/94}
\vspace{0.8cm}
\centerline{\tenrm A. BOTTINO, N. FORNENGO, G. MIGNOLA}
\baselineskip=13pt
\centerline{\tenit Dipartimento di Fisica Teorica, Universit\`a di Torino and}
\baselineskip=12pt
\centerline{\tenit INFN, Sezione di Torino, via P.Giuria 1,
10125 Torino, Italy}
\vspace{0.3cm}
\centerline{\tenrm and}
\vspace{0.3cm}
\centerline{\tenrm S. SCOPEL}
\baselineskip=13pt
\centerline{\tenit Dipartimento di Fisica, Universit\`a di Genova and}
\baselineskip=12pt
\centerline{\tenit INFN, Sezione di Genova, via Dodecaneso 33,
16146 Genova, Italy}
\baselineskip=13pt
\vspace{0.9cm}
\centerline{\it presented by A. Bottino
\footnote{Invited Talk to
{\it Physics from Planck Scale to Electroweak Scale}, Warsaw,
September 1994.}}
\vspace{0.9cm}
\abstracts{
Theoretical expectations for direct and indirect searches for relic
neutralinos are presented. Complementarity among various investigation
means is discussed in connection with the values of the neutralino
relic abundance.
}

\vfil
\rm\baselineskip=14pt
\section{Introduction}

In the present report we discuss the theoretical expectations for
detection of relic neutralinos either by the direct method
({\it e.g.} by
measuring the nuclear recoil energy due to neutralino--nucleus elastic
scattering) or by detecting indirect signals due to
neutralino--neutralino annihilation occurring in the halo or inside
celestial bodies (Sun and Earth).

Evidence for the presence of these relic neutralinos would obviously
be of primary importance as a test of the standard Big Bang theory,
independently of the fact that these Susy relic particles would or
would not provide a large contribution to the total mass of the
Universe. Should the relic neutralinos contribute substantially to
the total density parameter $\Omega$, their detection would also
provide an extraordinary hint for a clarification of the long--standing
dark matter issue. In fact, in this case, relic neutralinos would
play a significant role as Cold Dark Matter (CDM) constituents.

In the following we adopt the usual assumptions that:
i) R--parity is conserved, ii) the
neutralino is the lightest supersymmetric particle (LSP).
Very convenient theoretical frameworks where dark matter neutralino
phenomenology may be easily studied are provided by the Minimal
Supersymmetric Standard Model (MSSM) and by its implementation in a
Supergravity theory (SUGRA) \cite{susy,kane,arnowitt}. Here MSSM
is meant to denote the minimal supersymmetric extension of the
standard model where sleptons and squarks are taken as degenerate, with
the exceptions for the stop particles \cite{arnowitt}.

The neutralino $(\chi)$ is defined as the lowest--mass linear combination
of photino, zino and higgsinos

\beq
\chi=a_1\tilde\gamma + a_2\tilde Z+a_3\tilde H_1^0
+a_4\tilde H_2^0
\eeq

\noindent
Here $\tilde \gamma$ and $\tilde Z$ are the fields obtained
from the original U(1) and SU(2) neutral gauginos,
$\tilde B$ and $\tilde W_3$,
by a rotation in terms of the Weinberg angle.

The neutralino mass $\mchi$ and the coefficients $a_i$
depend
on the parameters: $\mu$ (Higgs mixing parameter),
$M_1$, $M_2$ (masses of {$\bi$}  and of $\wi$, respectively)
and $\tan \beta = v_u/v_d$ ($v_u$ and $v_d$ are
the v.e.v.'s which give masses to up--type and down--type quarks).
It is customary to employ the
standard GUT relationship between $M_1$ and $M_2$:
$M_1=(5/3) \tan^2 \theta_W M_2 \simeq 0.5~M_2$. We use this assumption
here.

In the following for the parameters $M_2$ and $\mu$ \ we will consider
the ranges: \break
20 \gev $\leq M_2 \leq$ 6 \tev, 20 \gev $\leq |\mu| \leq$ 3 \tev.
$\rm \tan \beta$ will be taken at the representative value
$\rm \tan \beta = 8$.

For the evaluation of the neutralino relic abundance and of the event
rates for direct and indirect neutralino detections one has also to
assign values to the masses of a large number of particles, namely
to the Higgs bosons and to the Susy scalar partners of leptons and
quarks: sleptons ($\tilde l$) and squarks ($\tilde q$). In the MSSM
scheme we consider here
these values are assigned arbitrarily: a standard procedure
consists in assuming mass degeneracy both for sleptons and for squarks
except for the stop particles \cite{arnowitt}.
As for the neutral Higgs bosons we
recall that in the MSSM there are three neutral Higgs particles:
two CP--even bosons $h$ and $H$ (of masses $m_h$, $m_H$ with $m_H>m_h$)
and a CP--odd one $A$ (of mass $m_A$). Once a value for one of these
masses (say, $m_h$) is assigned, the other two masses ($m_A$, $m_H$) are
derived through mass relationships depending on radiative effects.

Implementation of MSSM with supergravity sets a much more constrained
phenomenological framework, since SUGRA establishes strict relations
between all the masses in play and the few fundamental theoretical
parameters: $A$ and $B$ (appearing in the soft symmetry--breaking interaction
terms), $m_0$ (common scalar mass at the GUT scale), $m_{1/2}$ (common
gaugino mass at the GUT scale) and $\mu$. Furthermore, other specific
theoretical requirements (features of the symmetry breaking, condition
that the neutralino be the LSP, ...) strongly restrict the whole
parameter space. This has important consequences; for instance, it
constrains the neutralino to compositions with dominance of the gaugino
components. It has to be noted that important
constraints on the neutralino parameter space may be inferred from
the recent data on $b\rightarrow s + \gamma$ process \cite{btosgamma}.
The precise nature of these constraints still requires further investigation,
especially in view of the large uncertainties due to QCD--effects
\cite{QCD:btosgamma}.

In the present note the emphasis is on the
investigation power of various detection methods and
on their complementarity. On purpose, the model adopted here for
the calculations is as simple as possible (not too much constrained
by theoretical requirements) with choices for the free masses that are
only restricted by experimental bounds.

Our main concern is to discuss the minimal sensitivity
required in experimental devices in order to
undertake a significant investigation of neutralino dark matter. For
this reason we present here evaluations where the smallest values
compatible with experimental lower bounds are assigned to
the unknown masses; this usually provides maximal values for the signals.
To be definite, in
the following we will set the sfermion masses at the value
$m_{\tilde f} = 1.2 ~ m_\chi$, when $m_\chi > 45 $ \gev,
$m_{\tilde f} = 45$~\gev~otherwise. Only the mass of the top squarks are
assigned a larger value of 1 \tev. The Higgs mass $m_h$ is set at the value
of 50 \gev. The top mass is fixed at $m_t = 170$~\gev.

\section{Neutralino Relic Abundance}

For the computation of the direct and indirect event rates for
neutralino one has to use a specific value for the neutralino
density $\rho_{\chi}$. Obviously, it would be inappropriate to assign
to the neutralino local density $\rho_{\chi}$ the standard value for the
total dark matter density
$\rho_l = {\rm 0.3~\gev~\cm^{-3}}$,
 unless one specifically verifies that the neutralino relic abundance
$ \Omega_\chi h^2$ turns out  to be at the level of an
$(\Omega h^2)_{\rm min}$
consistent with $\rho_l$. This is why a correct evaluation of the event
rates for ${\chi}$ detection also requires a calculation of its relic
abundance.

Thus we evaluate $\Omega_\chi h^2$  and we determine $\rho_{\chi}$ by
adopting a standard \mbox{procedure \cite{[4]}}: when
$\Omega_\chi h^2 \geq (\Omega h^2)_{\rm min}$, we put
$\rho_\chi=\rho_l$; when $\Omega_\chi h^2$ turns out to be less than
$(\Omega h^2)_{\rm min}$, we take

\beq
\rho_\chi = \rho_l {\Omega_\chi h^2 \over (\Omega h^2)_{\rm min}}~.
\eeq

\noindent
Here $(\Omega h^2)_{\rm min}$ is set equal to 0.03.

For the neutralino relic abundance $\Omega_\chi h^2$  we employ the
results of our previous work \cite{[5]}.
In Fig.1 we display regions of the $M_2,\mu$ plane which
are characterized by different values of $\Omega_\chi h^2$.
Also shown in this figure are the iso--mass curves (dashed lines)
and the iso--compositions curves (solid lines). Along an
iso--composition line the composition parameter $P$ defined as the
gaugino fractional weight, {\it i.e.} $P=a_1^2+a_2^2$, is kept fixed.
In this figure, as well as in the following ones, only results
for positive $\mu$ are displayed.

The results of Fig.1 can conveniently be reported in a
$\Omega_\chi h^2$ vs. $\mchi$ plot, for fixed values of $P$.
This is done in Fig.2 where $\Omega_\chi h^2$ is plotted as a
function of $\mchi$ for three representative neutralino compositions: i) a
gaugino--dominated composition ($P = 0.9$), ii) a composition of maximal
gaugino--higgsino mixing ($P = 0.5$), iii) a higgsino--dominated composition
($P = 0.1$). As expected, out of the three
compositions displayed in Fig.2, the gaugino--dominated state
provides the largest values of $ \Omega_\chi h^2$.
In order to have more
substantial values of $\Omega_\chi h^2$, one has to consider purer
gaugino compositions ($P \gsim 0.99$). The very
pronounced dips in the plots of Fig.2 are due to the s--poles in the
$\chi$--$\chi$ annihilation cross section (exchange of the $Z$ and of
the neutral Higgses). The sharp decrease at 80--90~\gev~is due
to the opening of the $W^+ W^-$ and $ZZ$ final states in the
$\chi$--$\chi$ annihilation. We remind that
$\Omega_\chi h^2 \propto (<\sigmav>_{\mbi{int}})^{-1}$,
where $<\sigmav>_{\mbi{int}}$ is the annihilation cross section
times relative velocity, averaged over the neutralino thermal
distribution, integrated from the freezing temperature down to
the present temperature.

\section{Indirect Detection at Neutrino Telescopes}

Let us turn now to the indirect search for neutralino dark matter
which can be performed by means of neutrino telescopes
\cite{[8],Venezia94,DFTT34}.
Neutralinos, if present in our galactic halo as dark matter components,
would be slowed down by elastic scattering off the nuclei of the celestial
bodies (Sun and Earth) and then gravitationally trapped inside them.
Due to the process of neutralino capture
these macroscopic bodies could accumulate neutralinos which would
subsequently annihilate in pairs. An important outcome of this
$\chi$--$\chi$ annihilation would be a steady flux of neutrinos from
these celestial bodies.

The differential neutrino flux at
a distance $d$ from the annihilation region is given by

\beq
{dN_{\nu}\over dE_{\nu}}={{{\Gamma}_A}\over 4\pi d^{2}}\sum_{F,f}
B^{(F)}_{\chi f}{dN_{f \nu}\over dE_{\nu}}
\eeq

\noindent
where ${\Gamma}_A$ is the annihilation rate and $F$ denotes the
$\chi$--$\chi$ annihilation final states which are:
1) fermion--antifermion pairs , 2) pairs of neutral and charged
Higgs bosons, 3) one gauge boson--one Higgs boson pairs, 4) pairs of
gauge bosons; $B^{(F)}_{\chi f}$ denotes the branching ratio into
the fermion $f$ (heavy quark or $\tau$ lepton), in the channel $F$;
$dN_{f \nu}/dE_{\nu}$ denotes the differential distribution
of the neutrinos generated by the semileptonic decays
of the fermion $f$. The $\nu_{\mu}$'s,
crossing the Earth, would convert into muons
and generate a signal of up--going muons inside a neutrino
telescope. Calculations of this muon flux from the original
neutrino flux may be performed using standard procedures
\cite{[8],Venezia94,DFTT34}.

Particular care has to be taken in the evaluation of the annihilation
rate ${\Gamma}_A$. This quantity is given by \cite{[9]}

\beq
\Gamma_A={C\over 2} \tanh^2 \left({t\over \tau_A}\right)
\eeq

\noindent
where $t$ is the age of the macroscopic body ($t= 4.5~{\rm Gyr}$ for Sun and
Earth),
$\tau_A = (C C_A)^{-1/2}$, $C$
is the capture rate of neutralinos in the macroscopic
body and $C_A$ is the annihilation rate per effective volume of the
body. The capture rate $C$ is
provided by the formula \cite{[10]}

\beq
C={\rho_{\chi}\over v_{\chi}}\sum_{i}{\sigma_{{\rm el}, i}\over
m_{\chi}m_{i}}(M_{B}f_{i})\langle v^{2}_{esc}\rangle _i
X_i,
\eeq

\noindent
where $v_{\chi}$ is the neutralino mean
velocity, $\sigma_{{\rm el}, i}$ is the cross section of the neutralino
elastic scattering off the
nucleus $i$ of mass $m_{i}$
(for some properties of the elastic $\chi$--nucleus cross section
see next Sect.4 and Ref. 15),
$M_{B}f_{i}$ is the total mass of the element
$i$ in the body of mass $M_{B}$, $\langle v^{2}_{esc}\rangle_{i}$
is the square escape velocity averaged over the distribution of the
element $i$, $X_{i}$ is a factor which takes account of kinematical
properties occurring in the neutralino--nucleus interactions.
$C_A$ is given by \cite{[9]}

\beq
C_A={<\sigmav>_0 \over V_0} \left(
{\mchi \over {20~{\rm \gev}}}\right)^{3/2}
\eeq

\noindent
where $<\sigmav>_0$ is the annihilation cross section
times relative velocity, averaged over the neutralino thermal
distribution, at present temperature.
$V_0$ is defined as
$V_0=(3 m^2_{Pl} T / (2 \rho \times 10~{\rm \gev}))^{3/2}$
where $T$ and $\rho$ are the central temperature and the central
density of
the celestial body. For the Earth ($T=6000 ~\mb{K}$,
$\rho= 13~{\rm \gram} \cdot {\rm \cm}^{-3}$)
$V_0= 2.3 \times 10^{25} {\rm \cm}^3$,
for the Sun ($T=1.4 \times 10^7~\mb{K}$,
$\rho= 150~{\rm \gram} \cdot {\rm \cm}^{-3}$)
$V_0= 6.6 \times 10^{28}~{\rm \cm}^3$.

For the computation of the capture rate (and then also of $\tau_A$)
one has to use a specific
value for the neutralino density $\rho_{\chi}$ according to the
procedure explained in Sect.2.

The results of our evaluations for the quantities $C$, $C_A$ and $\Gamma_A$
are reported in Ref. 11,12.
{}From these results it turns out that:
i) for the Sun, the equilibrium between capture and annihilation is
reached over the whole $m_\chi$ range; ii) in the case of the Earth,
equilibrium is
not reached for $m_\chi \gsim m_W$, because of the substantial suppression
introduced in $\Gamma_A$ by the factor ${\rm \tanh}^2(t/\tau_A)$.

Now we report some of our results about the flux of the up--going muons
in the case of $\chi$--$\chi$ annihilation in the Earth.
In Fig.3 we show the fluxes of the
up--going muons as functions of $m_\chi$ for a number of values of the
neutralino composition $P$, for $\chi$--$\chi$ annihilation in the Earth.
The threshold for the muon energy is $E^{\rm th}_{\mu} = 2~\gev$.
We recall here that the present experimental upper bound
for signals coming from the Earth is
$4.0 \cdot 10^{-14} {\rm \cm}^{-2} {\second}^{-1}$ (90 \% C.L.)
\cite{[12]}.
By comparing this upper limit with our fluxes we see that
the regions explored by Kamiokande (at our representative point:
$\tan \beta = 8, m_h = 50~\gev$) concern the mass range
$50~\gev \lsim m_\chi \lsim 65~\gev$.
These regions are illustrated in Fig.4 in a $M_2$--$\mu$ plot. In this
figure we also display the regions which could be explored by a neutrino
telescope with an improvement factor of 10 (and of 100)
in sensitivity.

The location and the shape of the most easily explorable regions in the
$M_2$--$\mu$ plane depend on the Earth chemical composition and on the
neutralino composition in terms of the gaugino, higgsino components.
In fact the capture rate of neutralinos is more effective when
neutralino mass matches the mass of some of the main components
of the Earth and when the neutralino is a large gaugino--higgsino mixture.
Because of these two properties the signal is maximal along iso--mass lines
in the range 50--65~\gev, with elongations along iso--composition lines
of sizeable mixing.

In Fig.5 we report the fluxes
for up--going muons as functions of $m_\chi$
due to $\chi$--$\chi$
annihilation in the Sun.
As before the
threshold for the muon energy is $E^{\rm th}_{\mu} = 2~\gev$.
The evaluated fluxes are below the present experimental upper
limit of Kamiokande: $6.6 \cdot 10^{-14} {\rm \cm}^{-2} {\second}^{-1}$
(90 \% C.L.) \cite{[12]}.
The regions explorable by a neutrino
telescope with an improvement factor of 10 (and of 100)
in sensitivity respect to the present Kamiokande sensitivity
are displayed in Fig.6. Here the regions that are more easely
explorable, expand toward gaugino--dominated compositions since
spin--dependent cross sections (with exchange of light squarks)
are important in the capture of neutralinos by the Sun.

{}From these results it can be concluded that neutrino telescopes with an area
above $10^5~{\rm m}^2$ are very powerful tools for investigating
neutralino dark
matter in large regions of the parameter space. It also emerges from the
previous results that the signals from the Earth and from the Sun
somewhat complement each other to allow an exploration about DM
neutralino over a wide range of $m_\chi$.
In Ref. 12 has been derived the relation between the
exposure $At$ of a neutrino telescope ($A$ being the telescope
area, $t$ the live time) and the explorable range in the neutralino
parameter space, when the signal--to--background ratio is
optimized by appropriate angular selections.

\section{Direct Detection}

Another way to search for dark matter neutralinos is the
direct detection which relies on the measurement
of the recoil energy of nuclei of a detector, due to elastic scattering
of $\chi$'s. The relevant quantities to calculate are the differential rate
(in the nuclear recoil energy $E_r$):

\beq
{dR \over dE_r} = N_T {\rho_\chi \over m_\chi}
 \int^{v_{max}}_{v_{min}(E_r)} dv f(v) v
 {d\sigma_{el} \over dE_r} (v,E_r)
\eeq

\noindent
and the integrated rate $R_{\rm int}$, which is
the integral of Eq. (7) from the threshold energy
$E_r^{\rm th}$, which is a characteristic feature of the detector,
up to a maximal energy $E_r^{\rm max}$. In
Eq.(7) $N_T$ denotes
the number of target nuclei,
$d \sigma_{\rm el} / d E_r$ is
the differential elastic cross section and $f(v)$
is the distribution of
$\chi$ velocities in the Galaxy. It is important to
note again that the local density $\rho_\chi$ is evaluated here
according to the procedure discussed in Sect.2.
In general, the $\chi$--nucleus cross section has two contributions:
a coherent contribution, depending on $A^2$
($A$ is the mass number of the nucleus) which is
due to Higgs and $\tilde q$ exchange diagrams;
a spin--dependent contribution, arising from $Z$ and
$\tilde q$ exchange, proportional to
$\lambda^2 J(J+1)$.

By way of example, let us remind the expression of the coherent
cross section due to the Higgs--exchange \cite{[14]}:

\beq
\sigma_{{\rm el}, H} = {8 G_F^2 \over \pi} \alpha_H^2
A_i^2 {m_Z^2 \over m_h^4}
{m_i^2 m_\chi^2 \over {(m_i+m_\chi)^2}}
\eeq

\noindent
where $\alpha_H$ is a quantity depending on the neutralino--Higgs
and the Higgs--quarks couplings. It is worth mentioning that
$\alpha_H$ depends rather sensitively on the {$\chi$--composition} and
on a number of parameters, such as $\tb$ and the Higgs masses.

Except for very special points in the parameter space, the coherent
contribution to elastic cross section strongly dominates
over the spin--dependent one. For a detailed analysis on the
calculation of the
direct event rates see Ref. 19 and references quoted
therein. For an experimental overview about dark matter detectors
see Ref. 20.

Here, as an example, we simply report
in Fig.7 the event rates $R_{\rm int}$ for a Germanium detector as a function
of $m_\chi$ for neutralino compositions $P=0.1,0.5,0.9$. Rates are calculated
by integrating the differential rate of Eq.(7) over the electron--equivalent
energy range (2--4) KeV. In Fig.8 we show the regions of the
$M_2$--$\mu$ parameter space
which can be explored with an improvement of one and two orders
of magnitude in the sensitivity of the detector.

As for the shape of these regions we refer to the comments presented
above, in Sect.3, in connection with Fig.4. Again, the signals
are higher along the iso--mass line with an $m_\chi$ close to the mass
of the nuclei composing the detector. Thus, using detectors of
different compositions allows explorations of the $M_2$--$\mu$ plane
over a wide range in $m_\chi$. For instance, investigation of
regions with small $m_\chi$ values (of order of 10~\gev) with very low
threshold detectors \cite{[17]} would be very interesting. In fact this
$\mchi$ range
(which is excluded by accelerator data only under a number of
assumptions) is out of reach for the indirect detection discussed
in the previous Section.

\section{Comparison between Direct and Indirect Signals}

Fig.s 3,5,7 provide a comparison between the capabilities of the direct
detection versus the detection by neutrino telescopes. It has to be
noted that these figures refer to an arbitrary representative point of
the parameter space. However, since both (direct and indirect) signals
depend mainly on a common quantity, i.e. the $\chi$--nucleus elastic
scattering, the relative size is approximately independent of the
variation of the model parameters. The main departures from this feature
occur in the case of the signal from the Earth in regions where
the factor $\tanh^2 (t / \tau_A)$ is small and for the signal from the
Sun in regions where spin--dependent effects in the capture rate are
dominant.
Fig.s 3,5,7 also display how the the theoretical predictions compare with
the present sensitivity for each individual method.
A comparison of the relative power of the two detection methods for
exploring the $M_2$--$\mu$ plane is provided by Fig.s 4,6,8.

\section{Conclusions}

In the previous sections we have examined two types of detection
methods for relic neutralinos, one based on direct measurement and
one on the detection of indirect signals due to $\chi$--$\chi$
annihilation in celestial bodies.  As we have seen by the explicit
expressions given in Sect.s 3--4, in both cases the detection rates are
proportional to the neutralino local density and to the elastic
neutralino--nucleus cross section:
$R \propto \rho_\chi \sigma_{el}$
(for simplicity, here and
in the following, we consider the neutralino mass fixed at some
arbitrary value $m_{\chi}$; the other free parameters may vary in the
parameter space along iso--mass curves).

Because of the properties discussed in Sect.2, once the neutralino
relic abundance is evaluated, one has to distinguish between two cases:

\medskip
\begin{tabular}{ccccc}

\ \ & a) & $\Omega_\chi \geq \Omega_{min}$ & which entails &
$R \propto \sigma_{el}$    \\
\ \ & b) & $\Omega_\chi < \Omega_{min}$ & which entails &
$R \propto \sigma_{el}/<\sigmav>_{\mbi{int}}$    \\
\end{tabular}
\medskip

Now, if the strength of the couplings increases, and hence
$\sigma_{el}$ and $\sigma_{\mbi{ann}}$ increase, it turns out that also in
case b) the rates $R$ usually increase even if at a much less extent
than in case a). It then happens that for the two detection methods
discussed so far, the largest values of the detection rates for relic
neutralinos occur for models where $\tan\beta$ is large and
neutralino is largely mixed (this last condition has to be somewhat
relaxed for the signal from the Sun, as discussed above).
For these models, because of the
large value of the annihilation cross section, the neutralino relic
density is small. This is a scenario where the detection methods
under discussion have good chances to detect relic neutralinos, even
if these particles cannot provide a substantial contribution to the total
$\Omega$ of the Universe. Regions of the parameter space where this
situation occurs are defined by some authors as
cosmologically uninteresting and consequently disregarded. On the
contrary, we believe that they deserve a careful investigation,
since much interesting physics could occur there. The relevance of
this point is further confirmed by some recent results derived from
SUGRA theories where some of the universality properties, usually
required at the GUT scale, are relaxed \cite{Olech}. In fact it
turns out that, in this case, very interesting schemes with large
neutralino mixing, large $\tan\beta$ and small relic abundance naturally
emerge.

As we have seen, when $\Omega_\chi$ is large the two detection techniques
previously discussed have less chances to succeed (except for
the signal from the Sun, which however requires neutrino
telescopes of very large area).
In order to cover this case adequately, one may
resort to different detection methods. These are based on the
measurements of indirect signals due to neutralino--neutralino
annihilation in the Galactic halo. These have been widely discussed in
the literature; we refer to Ref. 23
for previous references and for a new
analysis aimed at the evaluation of the antiproton to proton ratio
$(\bar p / p)$ in cosmic rays. A common feature
of this class of measurements is that the event rates are typically
proportional to the square of the local neutralino density times the
annihilation cross section:
$R \propto \rho_\chi^2 <\sigmav>_0$. This implies
that, in this case, detection rates are high in regions of the
parameter space where the neutralino relic abundance is large.
By way of example, we give in Fig.9 the regions of $M_2$--$\mu$
plane that can be explored by measuring the $(\bar p / p)$ ratio
\cite{DFTT35}.

In conclusion we may state that complementarity among various
detection methods may potentially offer a good coverage of the
neutralino parameter space.


\vspace{0.35cm}
\section{References}
\vspace{-0.35cm}

\begin{thebibliography}{99}
\bibitem{susy}
H.P. Nilles, {\it Phys. Rep.} {\bf 110} (1984) 1; \hfill\break
H.E. Haber and G.L. Kane, {\it Phys. Rep.} {\bf 117} (1985) 75.

\bibitem{kane}
G.L. Kane, C. Kolda, L. Roszkowski and J.D. Wells, UM--TH--93--24 preprint;
\hfill \break
V. Barger, M.S. Berger and P. Ohmann, MAD/PH/801 preprint.

\bibitem{arnowitt}
R. Arnowitt and P. Nath, CTP--TAMU--52/93 preprint.

\bibitem{btosgamma}
J. Wu, R. Arnowitt and P. Nath, CTP--TAMU--03/94 preprint; \br
F.M. Borzumati, M. Drees and M.M. Nojiri, KEK--TH--400 preprint; \br
V. Barger, M.S. Berger, P. Ohmann and R.J.N. Phillips, MAD--PH--842
preprint.

\bibitem{QCD:btosgamma}
A. Ali and C. Greub, {\it Z. Phys.} {\bf C60} (1993) 433; \br
A.J. Buras, M. Misiak, M.M\"unz and S. Pokorski, MPI--PH/93--77
preprint.

\bibitem{[4]}
T.K. Gaisser, G. Steigman and S. Tilav, {\it Phys. Rev.}
 {\bf D34} (1986) 2206.

\bibitem{[5]}
A. Bottino, V.de Alfaro, N. Fornengo, G. Mignola and M. Pignone,
{\it Astroparticle Physics}
{\bf 2} (1994) 67. References to other recent
evaluations of the neutralino relic abundance are given in \cite{[6],[7]}.

\bibitem{[6]}
K. Griest, M. Kamionkowski and M.S. Turner,
{\it Phys. Rev.} {\bf D41} (1990) 3565; \hfill \break
J. McDonald, K.A. Olive and M. Srednicki,
{\it Phys. Lett.} {\bf B283} (1992) 80.

\bibitem{[7]}
P. Gondolo, M. Olechowski and S. Pokorski,
MPI--PH/92--81 preprint; \hfill \break
M. Drees and M.M. Nojiri, {\it Phys. Rev.}
 {\bf D47} (1993) 376; \hfill \break
P. Nath and R. Arnowitt, CTP--TAMU--66/92 preprint; \hfill \break
J.L. Lopez, D.V. Nanopoulos, and K. Yuan,
CTP--TAMU--14/93 preprint; \hfill \break
R.G. Roberts and L. Roszkowski, {\it Phys. Lett. }
{\bf B309} (1993) 329.

\bibitem{[8]}
G.F. Giudice and E. Roulet, {\it Nucl. Phys.} {\bf B316}
(1989) 429; \hfill \break
G.B. Gelmini, P. Gondolo and E. Roulet, {\it Nucl. Phys.}
{\bf B351} (1991) 623; \hfill \break
H. Kamionkowski,  {\it Phys. Rev.} {\bf D44} (1991) 3021; \hfill \break
A. Bottino, V. de Alfaro, N. Fornengo, G. Mignola and
M. Pignone, {\it Phys. Lett.} {\bf B265} (1991) 57; \hfill \break
M. Mori et al., {\it Phys. Rev.} {\bf D48} (1993) 5505; \br
M. Drees, G. Jungman, M. Kamionkowski and M.M. Nojiri,
{\it Phys. Rev.} {\bf D49}
(1994) 636; \hfill\break
R. Gandhi, J.L. Lopez, D.V. Nanopoulos, K. Yuan and A. Zichichi,
CERN--TH 6999/93.

\bibitem{Venezia94}
A. Bottino, N. Fornengo, G. Mignola and S. Scopel,
Proc. of the International Workshop on "Neutrino Telescopes",
Venice 1994.

\bibitem{DFTT34}
A. Bottino, N. Fornengo, G. Mignola and L. Moscoso,
DFTT 34/94 preprint (to appear in {\it Astroparticle Physics}).

\bibitem{[9]}
K. Griest and D. Seckel,  {\it Nucl. Phys.} {\bf B283} (1987) 681.

\bibitem{[10]}
A. Gould, {\it Ap. J.} {\bf 321} (1987) 571;
{\it Ap. J.} {\bf 328} (1988) 919;
{\it Ap. J.} {\bf 368} (1991) 610;

\bibitem{[11]}
A. Bottino, V. de Alfaro, N. Fornengo, A. Morales, J. Puimedon and
S. Scopel, {\it Mod. Phys. Lett.} {\bf A7} (1992) 733. \hfill \break
Notice that in the present note,
for the Higgs boson--nucleon coupling, we use
the evaluation of
J. Gasser, H. Leutwyler and M. E. Sainio,
{\it Phys. Lett.} {\bf B253} (1991) 252.

\bibitem{[12]}
M. Mori et al. (Kamiokande coll.),
{\it Phys. Lett.} {\bf B289} (1992) 463.

\bibitem{[13]}
J.G. Learned, {\it Nucl. Phys.} (Proc. Suppl.)
{\bf B31} (1993) 456 (Proc. Neutrino 92,
Ed. A. Morales); \hfill \break
F. Halzen, Proc.
{\it Fourth International Workshop on Neutrino Telescopes}
(1992), Ed. M. Baldo Ceolin; \hfill \break
L. Resvanis, Proc.
{\it Fifth International Workshop on Neutrino Telescopes}
(1993), Ed. M. Baldo Ceolin.

\bibitem{[14]}
R. Barbieri, M. Frigeni and G.F. Giudice,
{\it Nucl. Phys.} {\bf B313} (1989) 725.

\bibitem{[15]}
A. Bottino, V.de Alfaro, N. Fornengo, G. Mignola and S. Scopel,
{\it Astroparticle Physics} {\bf 2} (1994) 77.

\bibitem{[16]}
L. Mosca, Proc. of the XIVth Moriond Workshop {\it Particle
Astrophysics, Atomic Physics and Gravitation}, to appear; \br
See also the
Proceedings of the Workshop on
{\it The Dark Side of the Universe}, Rome, June 1993,
(Ed.s R. Bernabei and C. Tao) for extensive literature
about experiments of direct detection.

\bibitem{[17]}
S. Cooper et al., proposal of Max--Planck--Institut f\"ur Physik and Technische
Universit\"at M\"unchen, MPI--PhE/93--29.

\bibitem{Olech}
M. Olechowski and S. Pokorski, MPI--PHT/94--40 preprint
and these Proceedings.

\bibitem{DFTT35}
A. Bottino, C. Favero, N. Fornengo and G. Mignola,
DFTT 35/94 preprint (to appear in {\it Astroparticle Physic}).

\end{thebibliography}
\end{document}